\documentclass[aps,pre,twocolumn,superscriptaddress]{revtex4-2}
\usepackage{xcolor}
\usepackage{amsmath}
\usepackage{cases}
\usepackage{graphicx}
\usepackage{hyperref}
\begin{document}


\title{Optimal information transmission in a sequential model for cell division}


\author{Krishna P. Ramachandran}
\affiliation{Department of Physics and Astronomy, University of Pittsburgh, Pittsburgh, Pennsylvania 15260 USA}

\author{Motasem ElGamel}
\affiliation{Department of Physics and Astronomy, University of Pittsburgh, Pittsburgh, Pennsylvania 15260 USA}
\affiliation{Department of Physics and Initiative in Theory and Modeling of Living Systems, Emory University, Atlanta, Georgia 30322, USA}

\author{Farshid Jafarpour}
\affiliation{Institute for Theoretical Physics, Utrecht University, 3584 CC Utrecht, Netherlands}

\author{Andrew Mugler}
\email{andrew.mugler@pitt.edu}
\affiliation{Department of Physics and Astronomy, University of Pittsburgh, Pittsburgh, Pennsylvania 15260 USA}



\begin{abstract}
In proliferating cell populations, adaptive changes to biochemical reactions can change a cell's division time, which in turn can change the population size. However, biochemical reactions are subject to noise, and therefore the conditions for optimal information transmission from the molecular to the population scale are poorly understood.
Here, we model cell proliferation as a Bellman-Harris branching process with age-dependent division times. We identify a class of division time distributions, built from a series of Markovian steps, for which the population size distribution at all times is hierarchically calculable. We use this feature to characterize the amount of influence that a given reaction step has on the population size via the mutual information.
We find that information transmission is optimal for a characteristic number of steps until division: too few and the population size is unpredictable; too many and any given step has vanishing influence on the population size.
Our work reveals the potential tradeoffs involved in adaptive decision making at the sub-cellular, cellular and population scales.
\end{abstract}


\maketitle

\section{Introduction}

Division and proliferation of cells is the basis for the establishment of microbial communities, the formation of multicellular organs and organisms, and many other processes in biology. For unicellular organisms such as bacteria or other prokaryotes, virtually every activity undertaken during their life cycles is in service of survival or promoting their capacity for growth and proliferation. There are several examples to support that in general, bacterial behaviors are evolutionarily selected to account for survival and reproduction \cite{kussell2005, rainey2011, nadell2008}.
	
Cells make decisions favorable to their reproductive capacity on timescales shorter than evolutionary ones, without requiring mutations that drastically alter their physiology. These adaptive decisions are encoded in biochemical reactions and the responses of these reactions to environmental changes. From the standpoint of proliferation, the most relevant reactions are those that lead to cell division. Before dividing, cells complete a series of milestone processes such as DNA replication and septum formation \cite{rothfield1997, coltharp2016}. In response to an environmental change, a cell might speed up or delay these processes, decreasing or increasing its division time, leading to a larger or smaller population size, respectively \cite{jonas2014divide}.

Biochemical reactions are subject to intrinsic molecular stochasticity \cite{raj2008nature, elowitz2002stochastic}. This stochasticity leads to variability in cell division times, which in turn leads to variability in population sizes.
This makes the mapping from biochemical adaptation to population growth not only multiscale but noisy. As a result, the features of biochemical programs that give optimal control over population size are poorly understood.

Optimizing a functional mapping in the presence of noise is the domain of information theory \cite{cover1999elements}. Information theoretic tools such as entropy and mutual information have been used successfully to understand or predict biological design principles across scales \cite{bialek2012biophysics}. An open question is what division protocol allows optimal information transmission between its constituent biochemical steps and the ultimate size of the population.

Here, we introduce a stochastic, sequential model for cell division, and we ask which properties maximize the flow of information from the molecular to the population scale. We analyze the model using the mathematical formalism of the Bellman-Harris branching process, which connects sub-cellular timing stochasticity to population size distributions. We identify a class of division time distributions that makes the model hierarchically solvable. This allows us to calculate the mutual information between one or more sub-cellular reaction rates and the population size at a given time. We find that the mutual information is maximized at a particular number of reaction steps, and we interpret this result in terms of the tradeoffs between population growth rate, sensitivity and noise. We discuss our conclusions in a larger context and suggest possible directions for future work.

\section{\label{sec_model} Model}

We begin with a stochastic model of cell division, which we then propagate to population growth. Using this multiscale model, we investigate the information transmission from subcellular processes to population size.

\subsection{Cell division}
Cell division is an age-structured process, meaning that the probability for a cell to divide is peaked at a certain age since its birth. The cell cycle also contains checkpoints that must be passed before division occurs. In eukaryotic cells, these checkpoints are explicit cell cycle stages, but even in prokaryotes, division does not occur before important milestones are completed, such as the copying of the chromosome and the formation of important organelles.

The simplest process that contains multiple checkpoints and leads to a peaked division time distribution is a series of Markovian steps \cite{yates2017}. Markovian means that each step takes an exponentially distributed amount of time. A checkpoint can then be viewed as an individual step or as a subset of steps, in which case the checkpoint itself would also be age-structured. The final step is division. This process leads to a peaked distribution of cell cycle times (a hypoexponential distribution) that can take a broad range of shapes depending on the rates of the individual steps.

This model of cell division does not capture distributions typical of highly parallel processes, which may be important in bacterial proliferation \cite{pugatch2015greedy}. Nor do we consider correlations between subsequent cell cycle times, which are important for homeostatic control of cell size \cite{amir2014cell, willis2017sizing, susman2018individuality, elgamel2023multigenerational, elgamel2024effects}. As we will see, these omissions keep our process tractable for questions about information transmission across scales, and the above features could be added in future work.

\begin{figure}
	\includegraphics[width = \linewidth]{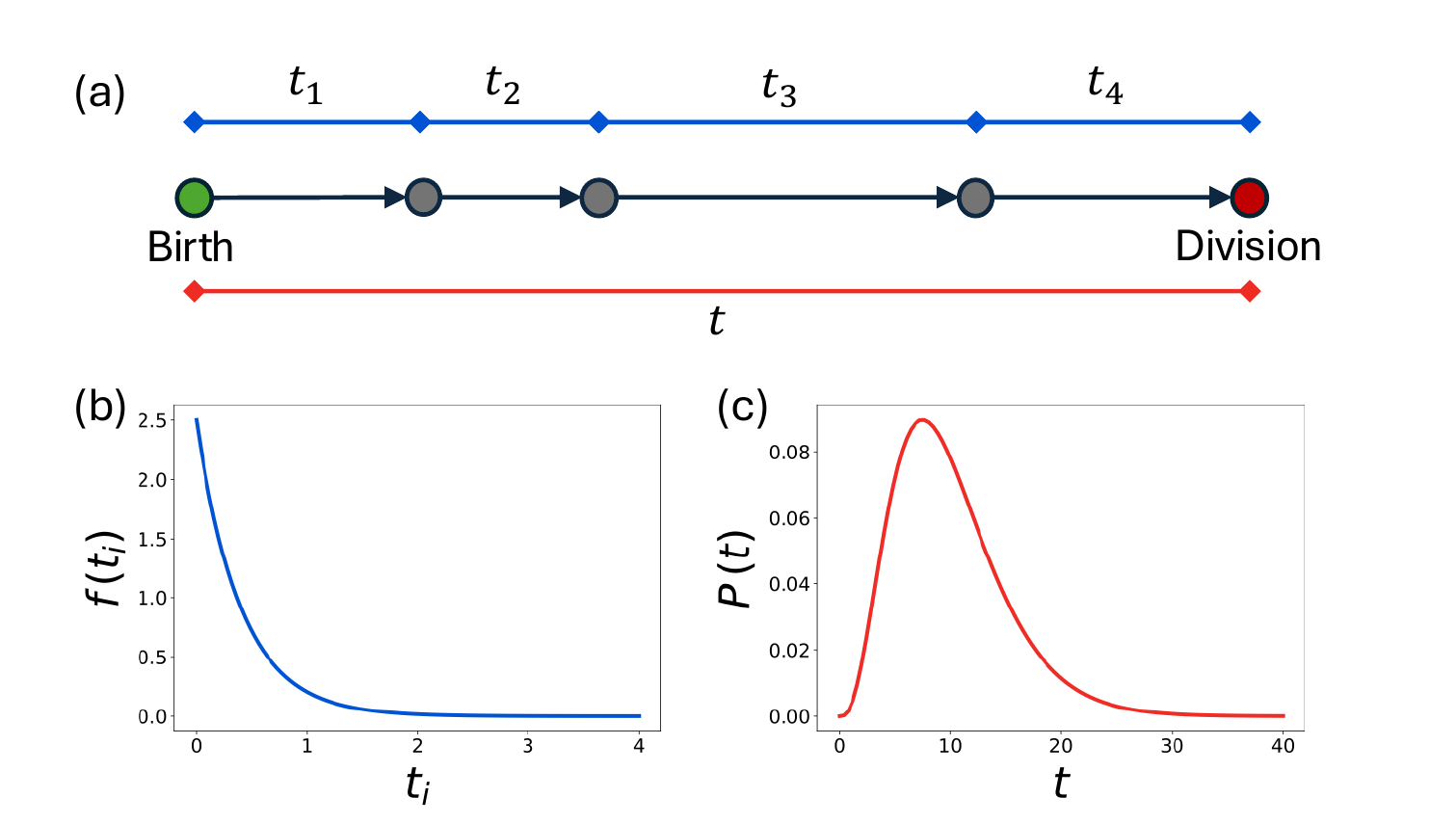}
	\caption{\label{fig1} (a) Representation of the sequential division process. (b) The times $t_i$ are drawn from exponential distributions for $i = 1, \dots, k$. (c) Consequently, the total birth-to-division duration $t$ is distributed hypoexponentially.}
\end{figure}

Consider a series of $k$ Markovian steps leading to cell division [Fig.\ \ref{fig1}(a)]. The duration $t_i$ of the $i$th step is exponentially distributed with rate $\gamma_i$ [Fig.\ \ref{fig1}(b)],
\begin{equation}
\label{f}
	f_i(t_i) = \gamma_i e^{-\gamma_i t_i}.
\end{equation}
The distribution of the total time $t$ from birth to division is then given by a convolution,
\begin{equation}
\label{P}
P(t) = \int_0^\infty dt_1\dots dt_k\ f_1(t_1)\dots f_k(t_k)\ \delta \left(t - \sum_{i=1}^k t_i \right),
\end{equation}
where the delta function enforces that the step durations sum to the total time. For unique rates, Eq.\ \ref{P} evaluates to a linear combination of the $k$ exponentials $f_i(t)$, which is known as a hypoexponential distribution. Degenerate rates lead to polynomial prefactors on the associated exponential. In general, this distribution will be peaked [Fig.\ \ref{fig1}(c)]. In the special case of all rates being equal, Eq.\ \ref{P} evaluates to the Gamma distribution, a peaked distribution that limits to a Gaussian as the number of steps gets large.

The key feature of Eqs.\ \ref{f} and \ref{P} that will make this model tractable for propagating to population growth is that $P(t)$ can be written as a linear combination of its derivatives. To see this, consider the case of $k=2$: Eqs.\ \ref{f} and \ref{P} read
\begin{equation}
\label{P2int}
P(t) = \int_0^t dt_1\ \gamma_1 e^{-\gamma_1t_1} \gamma_2 e^{-\gamma_2 (t-t_1)}.
\end{equation}
Differentiating Eq.\ \ref{P2int} gives
\begin{align}
\dot P &= \gamma_1\gamma_2e^{-\gamma_1t} + \int_0^t dt_1\ \gamma_1 e^{-\gamma_1t_1}\gamma_2(-\gamma_2) e^{-\gamma_2 (t-t')}  \nonumber\\
\label{P2inter}
&= \gamma_1\gamma_2e^{-\gamma_1t} - \gamma_2 P,
\end{align}
where the second step uses Eq.\ \ref{P2int} again. Differentiating Eq.\ \ref{P2inter} gives
\begin{align}
\ddot P &= (-\gamma_1)\gamma_1\gamma_2e^{-\gamma_1t} - \gamma_2 \dot P \nonumber \\
&= -\gamma_1(\dot P + \gamma_2P) - \gamma_2 \dot P,
\end{align}
where the second step uses Eq.\ \ref{P2inter} again. Rearranging gives
\begin{equation}
\label{Peq2}
0 = \ddot P + (\gamma_1 + \gamma_2)\dot P + \gamma_1\gamma_2P,
\end{equation}
and we see that $P(t)$ is a linear combination of its first two derivatives. In general, $P(t)$ is a linear combination of its first $k$ derivatives: the generalization of Eq.\ \ref{Peq2} for any $k$ is
\begin{align}
\label{diffeq_P}
0 =\ &A_0\partial_t^k P + A_1\partial_t^{k-1} P + A_2\partial_t^{k-2} P + \dots \nonumber \\
	&+ A_k P,
\end{align}
where
\begin{equation}
A_j \equiv \sum_{i_1<i_2<\dots<i_j}\gamma_{i_1}\gamma_{i_2}\dots\gamma_{i_j},
\end{equation}
or more explicitly, $A_0 = 1$, $A_1 = \sum_{i=1}^k \gamma_i$, $A_2 = \sum_{i=1}^k\sum_{j>i}^k \gamma_i \gamma_j$, $A_3 = \sum_{i=1}^k\sum_{j>i}^k\sum_{\ell>j}^k \gamma_i \gamma_j\gamma_\ell$, and so on, until $A_k = \prod_{i=1}^k \gamma_i$.
\subsection{Population growth}

\begin{figure}[bt]
	\includegraphics[width = \linewidth]{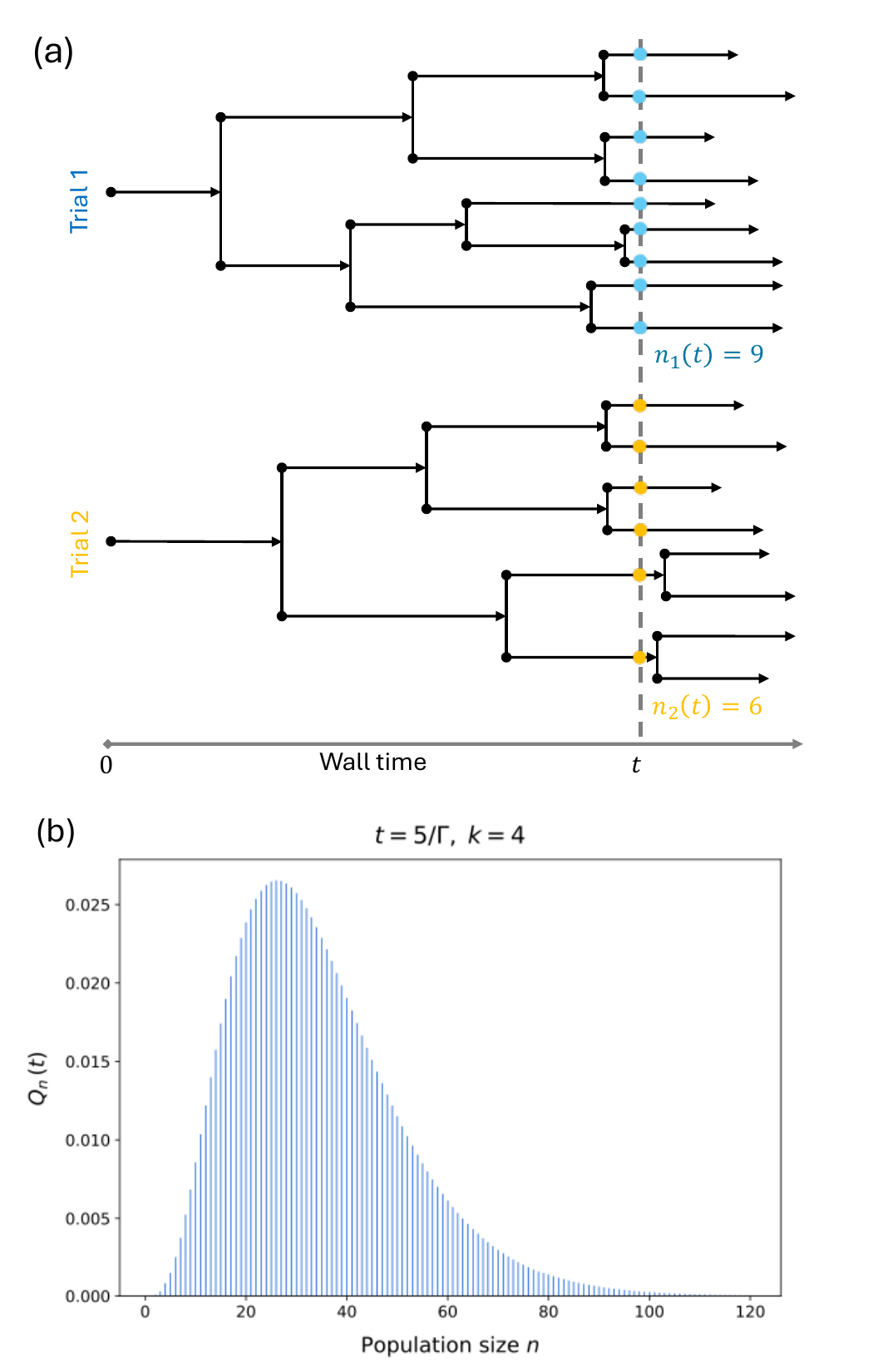}
	\caption{\label{fig2} (a) Schematic diagram of a branching process. Differing edge lengths denote differing ages at division. (b) Age heterogeneity leads to a distribution of population sizes at an given observation time across an ensemble of trials.}
\end{figure}

We consider a lineage of cells beginning with single cell that proliferates by binary division, drawing division times from the distribution $P(t)$. This is a stochastic branching process with two daughters at each branch. The variability in division times leads to a variability in population sizes $n$ across different realizations of the process [Fig.\ \ref{fig2}(a)]. We denote the distribution of population sizes at time $t$ by $Q_n(t)$ [Fig.\ \ref{fig2}(b)]. The corresponding probability generating function of $n$ is
\begin{equation}
\label{def_F}
F(z,t) = \sum_{n=1}^{\infty} z^n Q_n(t).
\end{equation}
The mathematical framework devised by Bellman and Harris for age-dependent branching processes \cite{bellman1948} connects the cell-scale timing variability to the population-scale number variability according to
\begin{equation}
\label{BH}
F(z,t) = z + \int_{0}^{t} dt' P(t-t') \left[F^2(z, t') - z\right],
\end{equation}
where the square reflects the binary division.

Equation \ref{BH} is a nonlinear integral equation for which there is no general solution. However, the fact that $P(t)$ is a linear combination of its derivatives allows us to turn Eq.\ \ref{BH} from an integral equation to a differential equation. Again, we illustrate for $k=2$. Differentiating Eq.\ \ref{BH} with respect to time twice gives
\begin{align}
\label{Fdot}
\dot F =\ &P(0)(F^2-z) + \int_{0}^{t} dt' \dot P(t-t') \left[F^2(z, t') - z\right], \\
\ddot F =\ &2P(0)F\dot F + \dot P(0)(F^2-z) \nonumber \\
\label{Fddot}
	& + \int_{0}^{t} dt' \ddot P(t-t') \left[F^2(z, t') - z\right].
\end{align}
Using Eq.\ \ref{Peq2}, we can write $\ddot P$ in Eq.\ \ref{Fddot} in terms of $\dot P$ and $P$. Then, we can use Eq.\ \ref{Fdot} to eliminate the integral with $\dot P$, and Eq.\ \ref{BH} to eliminate the integral with $P$. The result, after simplifying, is
\begin{equation}
\label{Feq2}
0 = \ddot F + (\gamma_1 + \gamma_2)\dot F + \gamma_1\gamma_2F(1-F),
\end{equation}
where we have used $P(0) = 0$ (Eq.\ \ref{P2int}) and $\dot P(0) = \gamma_1\gamma_2$ (Eq.\ \ref{P2inter}). Equation \ref{Feq2} is a second-order differential equation for $F$. Note that it has the same form as Eq.\ \ref{Peq2} but with $F\to F(1-F)$. In general, our class of $P$ gives a $k$th-order differential equation for $F$ with this property,
\begin{align}
\label{diffeq_F}
0 =\ &A_0\partial_t^k F + A_1\partial_t^{k-1} F + A_2\partial_t^{k-2} F + \dots \nonumber \\
	&+ A_k F(1-F)
\end{align}
(compare to Eq.\ \ref{diffeq_P}).

Inserting the definition of $F$ (Eq.\ \ref{def_F}) into Eq.\ \ref{diffeq_F} gives
\begin{align}
\label{insertQ}
0 =\ &A_0\sum_{n=1}^{\infty} z^n\partial_t^k Q_n + A_1\sum_{n=1}^{\infty} z^n\partial_t^{k-1} Q_n +
	\dots \nonumber \\
	&+ A_k \sum_{n=1}^{\infty} z^nQ_n\left(1-\sum_{m=1}^{\infty} z^mQ_m\right).
\end{align}
Equating coefficients of powers of $z$ gives
\begin{align}
\label{diffeq_Q}
0 =\ &A_0\partial_t^k Q_n + A_1\partial_t^{k-1} Q_n + A_2\partial_t^{k-2} Q_n + \dots \nonumber \\
	&+ A_k \left(Q_n - \sum_{m=1}^{n-1}Q_mQ_{n-m}\right).
\end{align}
Equation \ref{diffeq_Q} is a series of $k$th-order differential equations, one for each value of $n$. The series is hierarchical: the $n=1$ equation gives the dynamics of $Q_1$, the $n=2$ equation couples $Q_1$ to the dynamics of $Q_2$, the $n=3$ equation couples $Q_1$ and $Q_2$ to the dynamics of $Q_3$, and so on.

Equation \ref{diffeq_Q} is initialized using the conditions $Q_n(0) = \delta_{n1}$, and $\partial_t^i[Q_n]_{t=0} = 0$ for $i = 1, 2, \dots, k-1$. The first condition states that the population starts with one cell. The second condition comes from the fact that, for small $t$, a hypoexponential distribution scales as $P(t)\sim t^{k-1}$. Therefore, the probability that one or more cell divisions have occurred by time $t$ is $Q_{n>1}(t) = \int_0^t P(t') dt' \sim t^k$. Differentiating this quantity or its complement ($n=1$) between $1$ and $k-1$ times and evaluating at $t=0$ will give zero.

We solve Eq.\ \ref{diffeq_Q} numerically in hierarchical order by writing the equation for each $n$ as $k$ first-order equations. The result is an efficient computation method for $Q_n(t)$ at any time. See \cite{code} for the code.

\subsection{Information transmission}
Given an efficient way to compute $Q_n(t)$, we now ask about the mapping between the subcellular parameters $\{\gamma_i\}_{i=1}^k\equiv\vec{\gamma}$ and the population size $n$. Probabilistically, $Q_n$ is conditioned on the values of $\vec{\gamma}$, which we write as $Q(n|\vec{\gamma})$. The information contained in this conditional mapping is given by the mutual information \cite{cover1999elements, bialek2012biophysics}, defined
\begin{equation}
\label{I}
I(k,t) = \int d^k\gamma\ G(\vec{\gamma}) \sum_{n=1}^\infty Q(n|\vec{\gamma}) \log \frac{Q(n|\vec{\gamma})}{\int d^k\gamma' G(\vec{\gamma}') Q(n|\vec{\gamma}')},
\end{equation}
where $G(\vec{\gamma})$ is the distribution from which the parameters $\vec{\gamma}$ are drawn. Since $\vec{\gamma}$ and $n$ are integrated and summed over, respectively, $I$ is a function of the remaining dependencies: the number of steps $k$ and the observation time $t$.

The mutual information is a scalar that quantifies the dependence between two stochastic variables. If $Q(n|\vec{\gamma}) = Q(n)$, i.e., if the variables are independent, then $I=0$. Otherwise $I$ is positive, and its units depend on the base of the log (bits for $\log_2$, nats for $\ln$; here we use nats). Mutual information increases as the conditional distributions $Q(n|\vec\gamma)$ become more distinguishable across different values of $\vec\gamma$, i.e., as their overlap in $n$ decreases. This would represent more subcellular control of population size, as different settings of the subcellular parameters $\vec{\gamma}$ would lead to distinct population sizes $n$ with greater certainty.

$G(\vec{\gamma})$ determines which parameters can vary and how they are distributed. We consider two cases: (a) one rate can vary and (b) all rates can vary. These cases are given by
\begin{subnumcases}
{G(\vec{\gamma}) = }
	\label{Ga}
	{\cal U}(\gamma_1)\prod_{i=2}^k\delta(\gamma_i-\gamma), \\
	\label{Gb}
	\prod_{i=1}^k{\cal U}(\gamma_i),
\end{subnumcases}
where $\gamma$ is a constant, and ${\cal U}$ represents the uniform distribution from $0$ to $\gamma$. In Eq.\ \ref{Ga}, the first rate (chosen without loss of generality) varies uniformly while the others are fixed at $\gamma$ (later we will relax the assumption that $\gamma_1$ varies uniformly and explore other distributions). In Eq.\ \ref{Gb}, all rates vary uniformly and independently. In both cases, we allow $\gamma$ to scale with $k$ as $\gamma = \Gamma k$, so that the minimum mean division time (which occurs when all rates are $\gamma$) is a constant, $\Gamma^{-1}$, irrespective of $k$. We begin with the first case for simplicity, and we will see that our main result also holds in the second case.

Numerically, we approximate the integrals in Eq.\ \ref{I} as uniformly binned sums \cite{code}, and we find that above a certain number of bins our results do not change qualitatively. We choose observation times $t$ that are several multiples of $\Gamma^{-1}$, ensuring that many division events have occurred to generate the population. We will see that our main result is robust to this choice of observation time.

\section{\label{sec_results} Results}

\subsection{Optimal number of steps}
\begin{figure}
	\includegraphics[width = \linewidth]{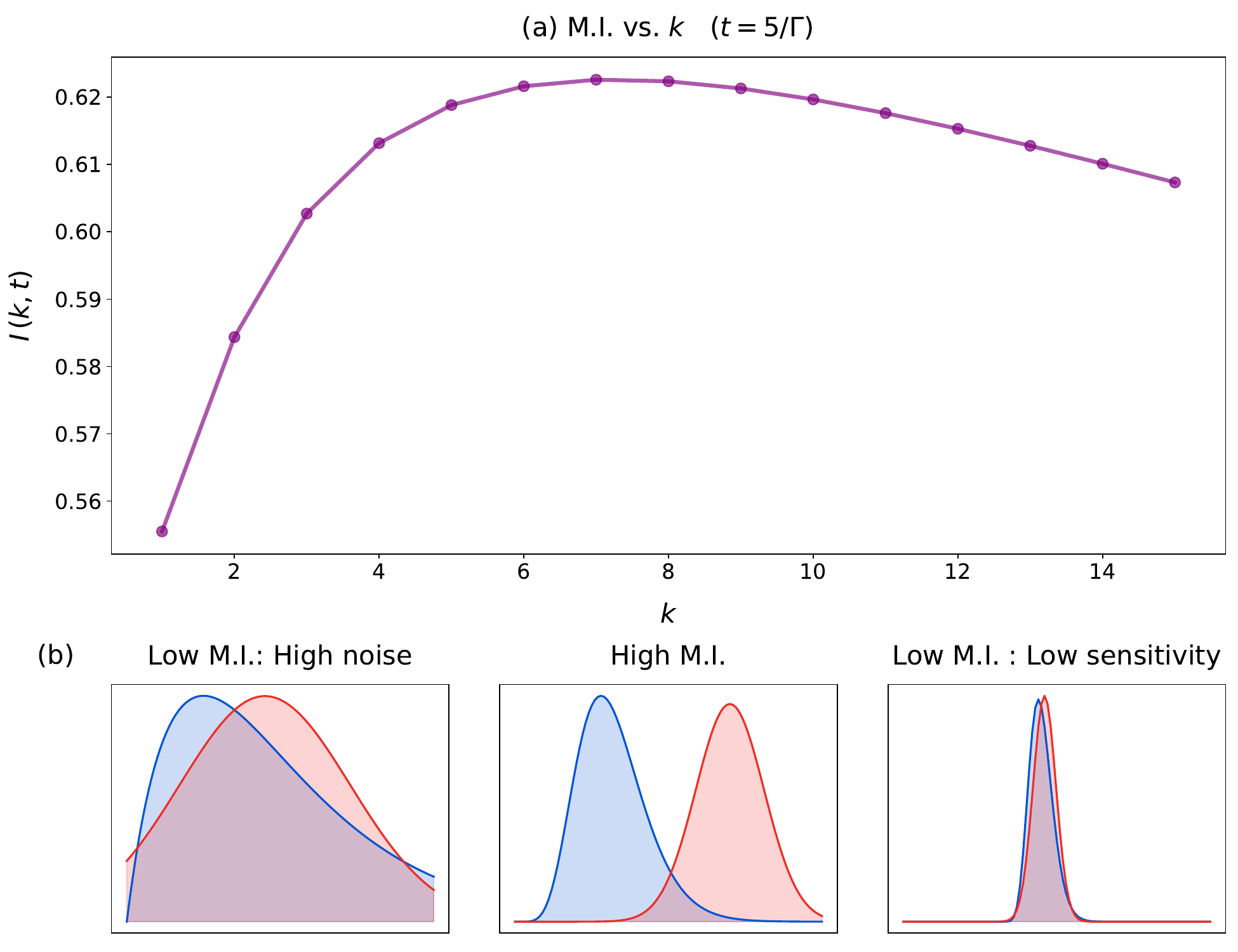}
	\caption{\label{fig3} (a) Mutual information $I$ between a subcellular rate and the population size, as a function of the number of subcellular steps $k$. We see that $I$ is maximized at an optimal $k_*$. (b) Schematic of population size distributions for two rate values at low (left), intermediate (middle), and high (right) $k$.}
\end{figure}

Figure \ref{fig3}(a) shows how mutual information varies as a function of the number of subcellular steps $k$, for observation time $T = 5/\Gamma$, when only one rate $\gamma_1$ varies (Eq.\ \ref{Ga}). We observe that there is a peak, in this instance at $k_*=7$. We will see that the peak can be understood as the result of a tradeoff between the uncertainty in the population size and the sensitivity of the population size to the rate $\gamma_1$. 

Qualitatively, we expect that low $k$ corresponds to a wider division time distribution $P(t)$. The reason is that, at high $k$, the randomness in individual steps averages out. Indeed, for the case where $P(t)$ is a Gamma distribution, the ratio of its standard deviation to its mean is $1/\sqrt{k}$, which vanishes for large $k$. Therefore, at low $k$, we might expect that a larger timing uncertainty is inherited as a larger population size uncertainty, widening $Q(n|\gamma_1)$ at low $k$. Wider distributions would have larger overlap for different choices of $\gamma_1$, reducing the information [Fig.\ \ref{fig3}(b), left].

On the other hand, high $k$ corresponds to many steps, where we expect any given step to have less of an impact on the completion time. Therefore, we might expect that at high $k$, the distribution $Q(n|\gamma_1)$ is less sensitive to $\gamma_1$. Lower sensitivity would also result in more overlap, which would also reduce the information [Fig.\ \ref{fig3}(b), right]. The optimal information at $k_*$ would then fall in between these two extremes, where the distributions are relatively narrow but still fairly separated [Fig.\ \ref{fig3}(b), middle].

\subsection{Uncertainty-sensitivity tradeoff}
To test the above expectations, we define measures of uncertainty and sensitivity, and we investigate how they vary with $k$. Specifically, we quantify sensitivity using the mean of $Q(n|\gamma_1)$, or $\bar{n}(\gamma_1)$. High sensitivity corresponds to a strong dependence of $\bar{n}$ on $\gamma_1$, or a large derivative $d\bar{n}/d\gamma_1$. Therefore, we define sensitivity as this derivative, scaled by $\gamma$, namely $d\bar{n}/d(\gamma_1/\gamma)$. We quantify uncertainty using the standard deviation of $Q(n|\gamma_1)$, or $\sigma_n(\gamma_1)$. High uncertainty corresponds to a large $\sigma_n$.

It turns out that for Gaussian variables, the ratio of these two quantities, $\rho \equiv [d\bar{n}/d(\gamma_1/\gamma)]/\sigma_n$, is directly related to the mutual information. Specifically, for sufficiently small $\sigma_n$, the mutual information is $\langle\log\rho\rangle$ up to an additive constant, where the average is over the distribution of $\gamma_1$ \cite{tkavcik2009optimizing}. Therefore, the quantity $\langle\log\rho\rangle$ will be useful for testing our expectations, if only approximately, since for us $\gamma_1$ and $n$ are not Gaussian, and $\sigma_n$ is not necessarily small.

\begin{figure}
	\includegraphics[width = \linewidth]{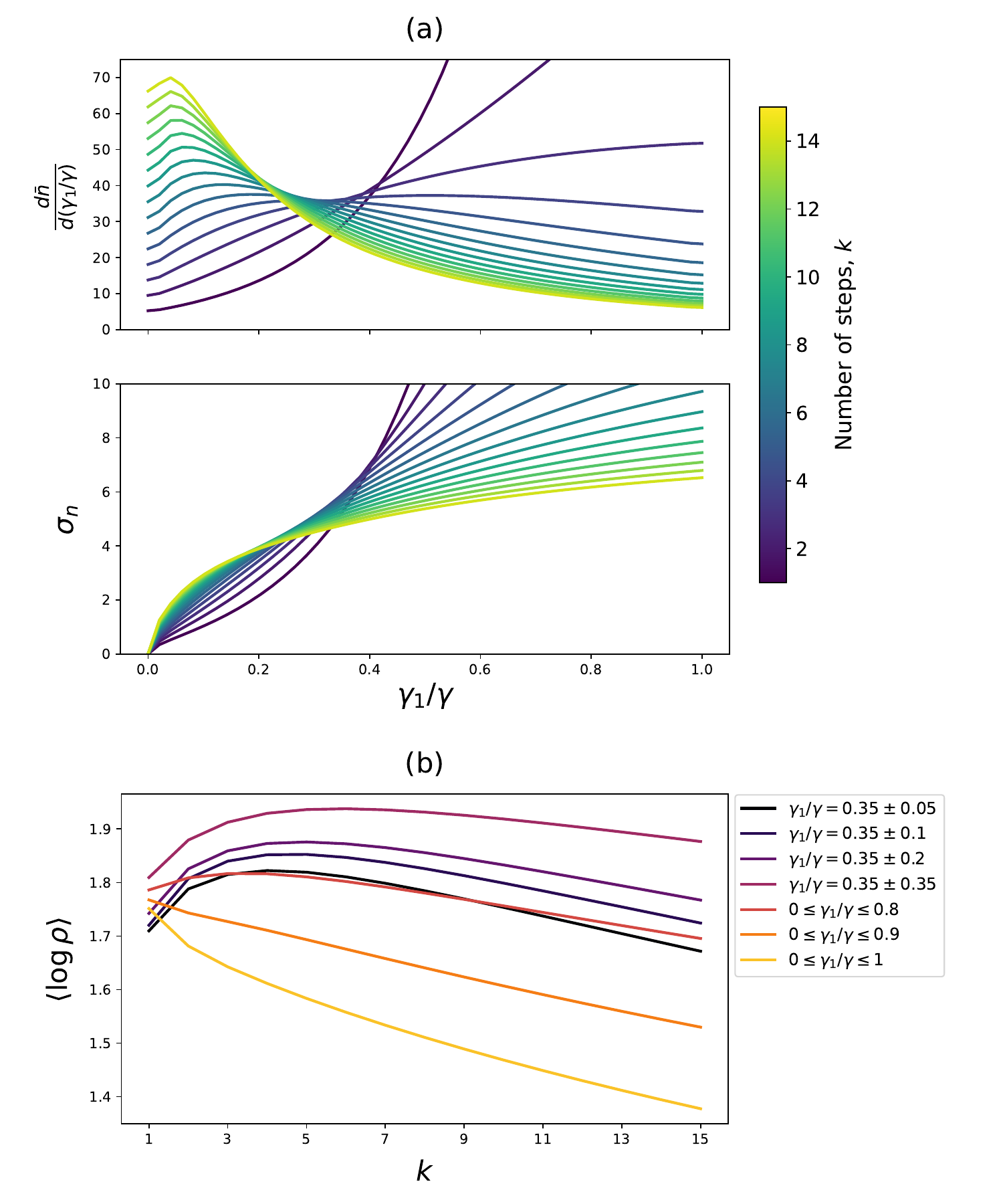}
	\caption{\label{fig4} (a) Sensitivity $d\bar{n}/d(\gamma_1/\gamma)$ (top) and uncertainty $\sigma_n$ (bottom) at various $k$ (color bar). For large $\gamma_1$, both decrease with $k$. (b) Their ratio, logged and averaged over $\gamma_1$, shows a peak with $k$. The average is taken over progressively larger uniform windows around $\gamma_1 = 0.35$ (color bar). Here $t=5/\Gamma$.}
\end{figure}

Figure \ref{fig4}(a) shows the sensitivity $d\bar{n}/d(\gamma_1/\gamma)$ (top) and uncertainty $\sigma_n$ (bottom) for various values of $k$. We see that, for large $\gamma_1$, both the sensitivity and the uncertainty decrease as $k$ increases (purple to yellow). This observation is consistent with our expectations that uncertainty is largest at low $k$ and sensitivity is smallest at high $k$. Interestingly, we see in Fig.\ \ref{fig4}(a) that for small $\gamma_1$, these dependencies are reversed, an observation we will return to in the next section when we investigate the effects of a nonuniform $\gamma_1$ distribution.

Figure \ref{fig4}(b) shows the ratio $\rho$ of sensitivity to uncertainty, logged and averaged over $\gamma_1$, as a function of $k$. We perform the average progressively (purple to yellow), steadily enlarging a uniform window around $\gamma_1/\gamma = 0.35$, the crossover location in Fig.\ \ref{fig4}(a). We see in Fig.\ \ref{fig4}(b) that, for all but the largest windows, $\langle\log\rho\rangle$ exhibits a peak in $k$, recovering the peak in the information seen in Fig.\ \ref{fig3}(a) and providing quantitative validation that the peak is due to a tradeoff between sensitivity and uncertainty. We attribute the loss of the peak when using the entire range of $\gamma_1$ (yellow) to the aforementioned facts that $\gamma_1$ and $n$ are not Gaussian and $\sigma_n$ is not necessarily small.

Altogether, we conclude that the tradeoff between maintaining high sensitivity and low uncertainty leads to an optimal number of steps $k_*$ that maximizes the information transmission between the subcellular rate $\gamma_1$ and the population size $n$.

\subsection{Robustness of findings}
\begin{figure}[h!]
	\includegraphics[width=\linewidth]{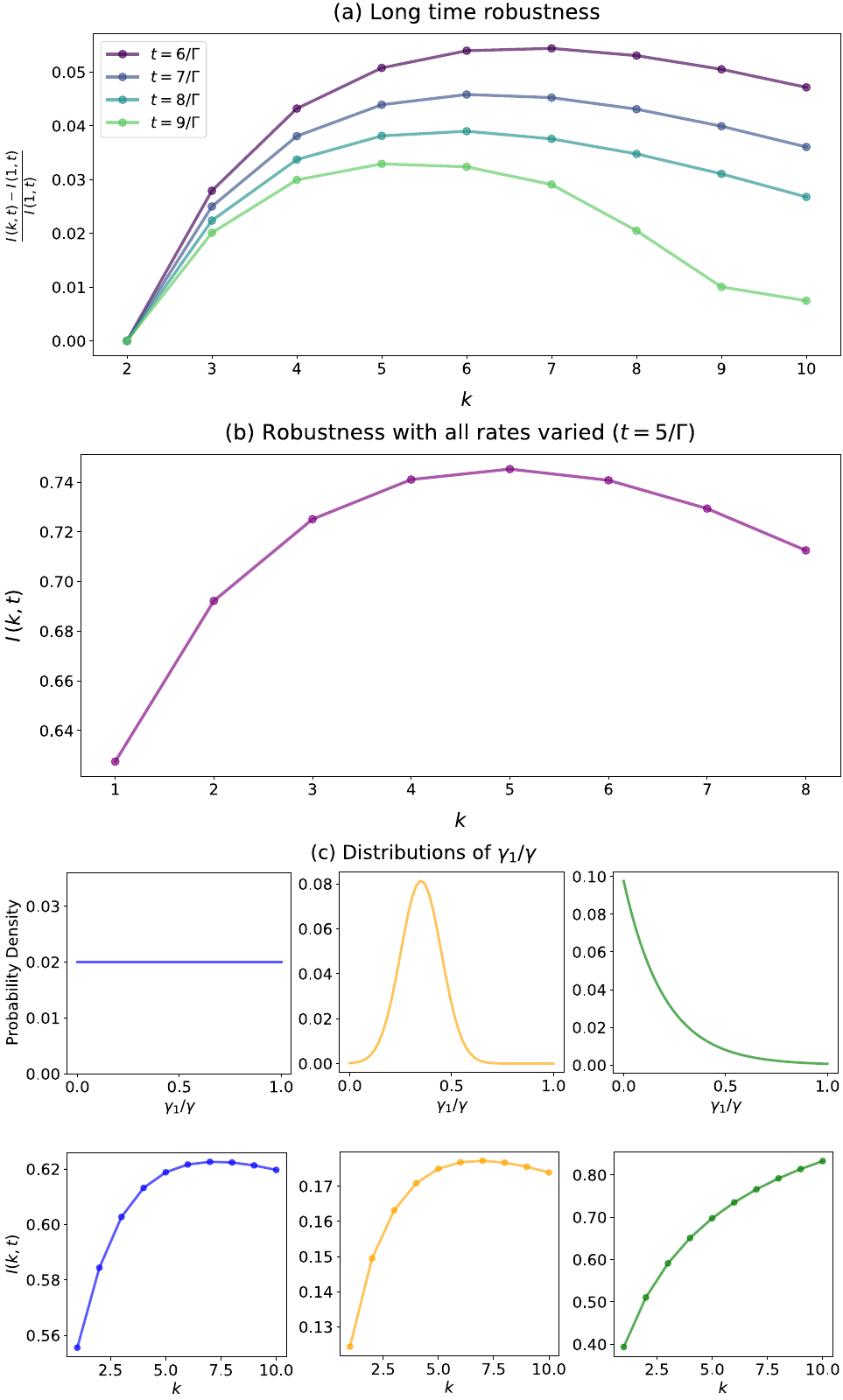}
	\caption{\label{fig5} Mutual information remains peaked (a) regardless of observation time $t$ and (b) when all rates are allowed to vary (Eq.\ \ref{Gb}). (c) The peak is also robust to the distribution of $\gamma_1$ except if $\gamma_1$ is restricted to the lower end of its range.}
\end{figure}

Finally, we test whether our finding of an optimal number of steps $k_*$ that maximizes information transmission $I$ is robust to (i) the observation time and (ii) how many rates are varied, and (iii) the distribution of the rate. Figure \ref{fig3}(a) was computed for a particular observation time of $t = 5/\Gamma$. Figure \ref{fig5}(a) shows that a maximum in $I$ as a function of $k$ persists, and even becomes more pronounced, as the observation time increases.

Figure \ref{fig3}(a) was also computed with only one rate $\gamma_1$ allowed to vary (Eq.\ \ref{Ga}). This choice was made for simplicity, but the more likely scenario is that a cell could vary more than one of its rates to change its division time and therefore the population size. In Fig.\ \ref{fig5}(b), we allow all rates to vary according to Eq.\ \ref{Gb}. Importantly, we see that the maximum in $I$ as a function of $k$ seen in Fig.\ \ref{fig3}(a) persists under this more general protocol.

Lastly, Fig.\ \ref{fig3}(a) was computed with a uniform distribution of $\gamma_1$, which we reproduce in Fig.\ \ref{fig5}(c) (left). In Fig.\ \ref{fig5}(c) (middle), we repeat with a Gaussian distribution and see that the maximum in $I$ as a function of $k$ persists. However, in Fig.\ \ref{fig5}(c) (right), we repeat with an exponential distribution and see that $I$ instead increases monotonically with $k$. The key difference with an exponential distribution is that it is biased toward smaller values of $\gamma_1$. Small $\gamma_1$ is the regime in which the dependencies of sensitivity and uncertainty on $k$ reversed [Fig.\ \ref{fig4}(a)]. In this regime, large $k$ corresponds to somewhat larger uncertainty but significantly larger sensitivity, making large $k$ advantageous for information transmission. Altogether, the results of Fig.\ \ref{fig5}(c) suggest that our main finding of an optimal number of subcellular steps is insensitive to the distribution of subcellular rates unless one rate ($\gamma_1$) is consistently lower than the others, i.e., it is strongly rate-limiting.

\section{Discussion}

We have proposed a simple stochastic model for cell division and used the mathematical pipeline provided by the Bellman-Harris branching process to investigate how the information encoded in the stochasticity of a molecule-scale process is transmitted to the population scale. Our numerical results indicate that information travels optimally across scales at a finite number of steps before division, illustrating the potential benefits for cells to use a checkpoint-based division mechanism from an optimal design perspective. More generally, our model serves as an example to demonstrate how to achieve optimal information transmission across scales in the presence of stochasticity.

Our model assumes that each checkpoint in the sequence leading to division is achieved in a memoryless fashion, its duration being drawn from an exponential distribution. In reality, not all sub-cellular checkpoints may be of a Markovian nature. However, if we interpret a checkpoint as comprising several of the steps, the checkpoint is no longer Markovian. Instead, the checkpoint is itself a hypoexponential distribution, providing more freedom to the interpretation of the model. Conversely, one could view the use of the exponential distribution for the duration of a checkpoint as a coarse-graining choice, replacing the time-varying instantaneous rate of completion of a general non-Markovian process with its average throughout the duration. Either way, it would be interesting to see if non-Markovian generalizations to our sequential model would yield any qualitative change to our main results.

A crucial assumption of our model is that the process leading up to division is serial. While this may apply to some cell types, cell division is, in general, preceded by simultaneous, parallel sub-cellular milestones \cite{pugatch2015greedy, micali2018}. The division-age distributions resulting from such parallel processes are less amenable to the branching process framework used in our analysis, as the unique characteristics of the hypoexponential distribution make $Q_n$ numerically computable. In the more general case, one would have to resort to stochastic simulations, which take a large number of trials to converge to the smooth population-size distributions necessary for computation of information-theoretic measures.

We consistently find that the number of effective subcellular steps that maximizes information transmission is on the order of $k_*\sim 5$$-$$10$. This finding has implications for the structure of the division time distribution $P(t)$. A distribution built from five to ten exponential steps is generally peaked but not extremely sharp, which is qualitatively consistent with experimentally measured division time distributions \cite{iyerbiswas2014}. Quantitatively, bacterial division time data are better described by heavier-tailed distributions, such as the log-Frechet distribution \cite{pugatch2015greedy}, rather than the Gamma distribution, which is a limiting case of our model. It remains an open question whether cell division time statistics can be comprehensively captured by a hypoexponential distribution as we study here. Certainly, there are some processes in the bacterial cell cycle that proceed in parallel \cite{haeusser2008, micali2018, colin2021} potentially with significant effects on division times.

In a broader sense, we have addressed system design principles that lead to optimal transmission of information in a multiscale proliferating stochastic system. The question remains as to what the evolutionary drive behind optimizing information flow may be. From an optimal control perspective, it is reasonable to hypothesize that cells as agents may compute long-term reward function during their growth and proliferation, in which information encoded in stochasticity could play a role. Future work on optimal control strategies at the cell level could shed more light on the relevance of information-theoretic measures to short- or long-term evolutionary outcomes.

\begin{acknowledgments}
This work was supported by National Science Foundation grant DMS-2245816 and National Institutes of Health grant R35GM156451. This work was performed in part at the Aspen Center for Physics, which is supported by National Science Foundation grant PHY-2210452.
\end{acknowledgments}

%

\end{document}